\documentclass{aa}

\usepackage{epsfig}
\usepackage{graphicx}

\makeatletter

\def\compoundrel#1\over#2{\mathpalette\compoundreL{{#1}\over{#2}}}
\def\compoundreL#1#2{\compoundREL#1#2}
\def\compoundREL#1#2\over#3{\mathrel
      {\vcenter{\hbox{$\m@th\buildrel{#1#2}\over{#1#3}$}}}}
\makeatother

\begin{document}
   \title{Scattering in the vicinity of relativistic jets:
          a method for constraining jet parameters}

   \author{C. K. Cramphorn$^{1,3}$, S. Yu. Sazonov$^{1,2}$
           \and R. A. Sunyaev$^{1,2}$}

   \offprints{conrad.cramphorn@vovo.de}
 
   \institute{$^1$Max-Planck-Institut f\"ur Astrophysik,
              Karl-Schwarzschild-Str.~1,
              85740 Garching bei M\"unchen, Germany\\
              $^2$Space Research Institute, Russian Academy of Sciences,
              Profsoyuznaya 84/32, 117997 Moscow, Russia\\
              $^3$Kanzlei Dr.~Volker Vossius,
              Geibelstr.~6,
              81679 M\"unchen, Germany}

   \date{Received 01/02/2004; accepted 11/03/2004}

   \abstract{Relativistic jets of radio loud active galactic nuclei (AGN)
             produce highly directed, intense beams of radiation.
             A fraction of this beamed radiation scatters
             on the thermal plasma generally surrounding an AGN.
             The morphology of the
             scattered emission can thus provide
             constraints on the
             physical properties of the jet. We present a model to study
             the feasibility of constraining the parameters of a jet,
             especially its inclination
             angle and bulk Lorentz factor
             in this way. We apply our model to the
             well studied jet of M87 and the surrounding diffuse gas and
             find that the observational limits of the surface brightness
             measured in the region of the putative counterjet provide
             the tightest constraints on the jet parameters consistent
             with constraints derived by other methods.
             We briefly discuss the applicability of our model to other
             sources exhibiting relativistic motions.
}

   \titlerunning{Scattering in the vicinity of relativistic jets}

   \maketitle

\section{Introduction}
Giant ellipticals
in the central parts of clusters of galaxies
often harbour an active galactic nucleus (AGN)
which is believed to be
powered by a supermassive black hole.
X-ray observations have revealed that these AGN
are surrounded by an atmosphere
consisting of a hot and tenuous thermal plasma
that fills the intracluster space.
The optical depth for Thomson scattering
due to the free electrons present in this plasma
reaches values of the order of
$10^{-3}-10^{-2}$. Therefore, up to one per cent of the
luminosity of an AGN radiating in such an environment
will be scattered on the free thermal electrons
and this scattered radiation will be
highly polarised (Sunyaev 1982).

A rather interesting situation arises in the case
of an AGN powering a radio galaxy through
a pair of relativistic jets which emit synchrotron radiation.
The radiation produced by the relativistic particles gyrating along
the magnetic fields of the jet is strongly beamed, i.e.~the radiation field
is highly anisotropic in the vicinity of the jets,
due to well known relativistic effects.

The free electrons present in the hot thermal plasma,
on the other hand, are distributed
in a more or less spherically symmetrical way around the AGN.
The electrons located close to the jet
are exposed to the largest radiative flux and, therefore,
will be the strongest sources of scattered radiation.
Thus, the surface brightness profile of the electron scattered
component should reflect the anisotropy of the jet emission and
could provide information about the intrinsic jet properties.
Combining this information with knowledge about the
density distribution of the scattering gas, which for instance is
obtainable through X-ray observations, might enable one
to gain information about such parameters as
the inclination angle of the jet axis towards the
line of sight and the bulk Lorentz factor of the jet flow.
The feasibility of this method will be studied in this paper.

The ``standard model'' of extragalactic jets
(Blandford \& Rees 1974; Scheuer 1974)
describes such objects as
intrinsically symmetric structures.
The different appearances of the jet pointing
towards the observer in comparison
with the one pointing away from the observer, the
so-called counterjet, stems primarily from the orientation of the
oppositely directed axes of these jets with respect to the observer.
In the case of a small inclination angle and
a large Lorentz factor the counterjet
might be rendered practically invisible, as is presumably the case in
e.g.~M87, offering only indirect evidence for its existence
in the form of e.g.~corresponding hotspots and radio lobes.

With respect to the question addressed in this paper, however,
it is important to note
that the electrons being illuminated by the counterjet
are exposed to and scatter as much radiation
as the electrons located on the side of the jet.
The only difference obviously being that the photons scattered
by electrons located on the side of the counterjet have to travel larger
distances in order to reach the distant observer
than photons scattered by electrons located on the side
of the jet.

The observational problem related to the detection of the discussed
effect against the ``background'' produced by
other sources of radio emission, e.g.~the
jets themselves, radio lobes etc.,
should be least demanding in the region of the counterjet.
In fact one can envisage a situation where the jet
pointing away from the observer is
beamed so strongly out of our direction
that the this jet is rendered practically invisible.
However, the scattered light produced by the counterjet might
still be visible and, thus, revealing its presence.

This idea was put forth by Gilfanov, Sunyaev \& Churazov
(1987, henceforth GSC87) in a study
of the scattered radiation originating from a point source,
which was modeled to vary over time
and to radiate non isotropically.
It was shown that a highly directional compact source
can produce extended jet-like
structures in the surface brightness distribution
of the scattered radiation. Furthermore, GSC87 demonstrated that
in addition to a temporal variability
a rotation as well as a precession
of the directional radiation diagram of the
central point source give rise to a large
amount of morphological variety of the diffuse sources of radiation.

In the context of similar studies
Wise \& Sarazin (1992) and Sarazin \& Wise (1993)
point out
that observations of electron scattered profiles
in cluster radio sources would provide a test of theories, which
seek to unify disparate types of AGN, viz.~the hypothesis that
Fanaroff-Riley Class I (FR I) radio galaxies are the parent population
of BL Lac objects.

The radiation beam produced by a
FR I radio source, which is pointing towards us in a BL Lac object, should
be observable by means of the scattered radiation it produces.

For the purposes of the above-cited studies it was appropriate
to model the radiation field produced by a relativistic jet
as two oppositely directed radiation
cones of constant intensity with a half-opening angle $\sim 1/\gamma$, where
$\gamma$ is the bulk Lorentz factor of the jet flow. Although
this approximation certainly is valid on scales much larger
than the jet itself (as will be confirmed by the results of this paper),
the effects caused by the exact shape of the beamed radiation pattern
as well as by the motion of the radiation source itself
should become noticeable as one ``approaches'' the jet,
i.e.~as one is interested in the behaviour of the scattered radiation
in the vicinity of the jet.
It is, therefore, one object of this paper
to study how well said approximation
works on length scales which are comparable to
the size of the jet itself.

This paper is structured as follows.
In Section 2 we describe
our model and the accompanying equations. We apply our
model to two scenarios in Section 3 and present
the numerically obtained solutions thereof.
We use these results in Section 4
to try to derive constraints on the inclination angle and
the bulk Lorentz factor for the observationally
well studied jet of M87.
Section 5 closes with a summary of our results.

\section{The model}
Radio observations have revealed the
complex structure of extragalactic jets (e.g.~Bridle \& Perley 1984).
Bright patches moving with apparent superluminal motion
often alternate with fainter regions along a given jet.
The origin of these bright patches, the so-called knots,
is still not fully understood. The most popular model at present,
the internal shock model (Rees 1978),
regards these knots as jet plasma excited by shocks
moving along the jet. Particles are accelerated by these shocks
and emit radiation by the synchrotron process as they gyrate
along the magnetic field of the jet. According to this model,
the velocity of a knot does not necessarily equal
the actual velocity of the underlying fluid flow, should,
however, represent an upper limit thereof.

\begin{figure}
\centering
\includegraphics[width=\columnwidth]{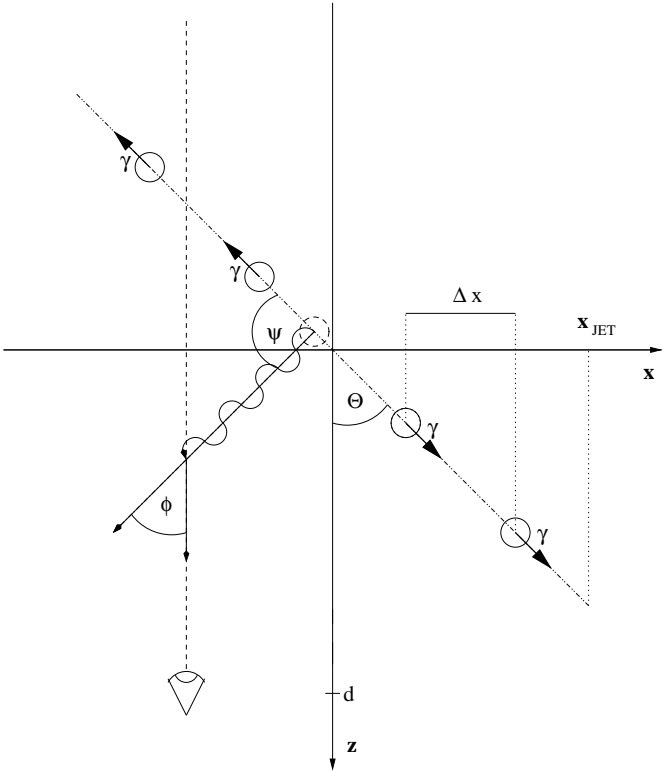}
 \caption{A sketch of the scenario studied in this paper.
          An AGN situated at the origin produces a two-sided,
          symmetric jet lying in the (x,z)-plane.
          The jet's radiation field is modeled by a chain of ``blobs'' ejected
          by the AGN at a constant rate along directions spanning an angle
          of $\theta$ and $\pi-\theta$ with the line of sight of the
          observer, who is located at a distance $d$ along the z-axis.
          These blobs (in this case two blobs on each side) travel with
          a relativistic velocity $\beta c$ corresponding to a
          Lorentz factor $\gamma$ along the respective jet axes.
          The blobs are separated by a projected physical distance
          $\Delta x$. The actual observed apparent distance between the blobs
          $\Delta x_\mathrm{app}$ differs for the two sides of the jet.
          After a blob has traveled a projected
          distance of $x_\mathrm{jet}$ away from the AGN,
          according to our model the blob ceases
          radiating.
          An exemplary photon (curly line), which was produced by one of the
          blobs at a time when the blob was closer to the nucleus
          (position marked by a dashed
          circle), is emitted along a
          direction making an angle $\psi$ with the direction
          of the blob motion along the counterjet. Further along its path the photon
          is scattered through an angle $\phi$
          into the direction of the distant observer
          and received by him at a distance $d$.
        }
\end{figure}

In another, somewhat less popular
model these knots are interpreted
as magnetised clouds of relativistic
electrons, so-called plasmoids or blobs,
being ejected out of the centre of an AGN
by a cataclysmic event
along the jet axes
(e.g. van der Laan 1963).
The velocity of these blobs by definition
corresponds to the velocity of the jet flow.

Irrespective of which of the above models is ``correct'',
in the following sections it is our aim to construct
a simple albeit realistic model of the radiation field produced by a
two-sided, symmetric relativistic jet. According to our model the jet
spans an inclination angle towards the line of sight $\theta$,
exhibits a bulk Lorentz factor $\gamma$ and radiates with
an intrinsic luminosity $L_\mathrm{jet}$.
We imagine the jet to be made up of several point sources moving along the
jet axis, which we will henceforth call ``blobs''.

We choose our system of coordinates as shown in Fig.~1.
The electron number density $n_\mathrm{e}(r)$ is
distributed in a spherically symmetrical way about the centre of the AGN.
The x- and y-axis span the plane of the sky, with the
x-axis being the projection of the jet axis
onto the plane of the sky.
The z-axis is pointing towards the observer.
The jet is inclined with respect to the line of sight
by an angle $\theta$ and the counterjet by an angle $\pi-\theta$,
respectively.

At a time $t = 0$ a first pair of blobs is created inside the
nucleus of the AGN.
Information about this event reaches the observer, located at
a distance $d$ along the z-axis, after a time $t = d/c$,
where $c$ is the speed of light.
The blobs are assumed to move ballistically with
a velocity $v=\beta c$ along the jet
and the counterjet, respectively, wherein $\beta$ is given by
$\beta=\sqrt{1-1/\gamma^{2}}$.
The motion of a blob can be described in the chosen coordinate system as
\begin{equation}
\textbf{r}_\mathrm{blob}(t)=\textbf{r}_\mathrm{o}+\textbf{v}t=
                        (x_\mathrm{o}+v_{x}t,\,
                         0,
                         z_\mathrm{o}+v_{z}t),
\end{equation}
with the projections of the blob velocity onto the coordinate axes
$v_{x}=v\sin{\theta}$
and $v_{z}=v\cos{\theta}$, respectively. For a single pair of blobs
$x_\mathrm{o}$ and $z_\mathrm{o}$ could be omitted.
However, these quantities become necessary once one wishes to keep
track of several pairs of blobs over time as described in the later sections
of this paper.

The apparent velocity $v_\mathrm{app}$
of the blobs in the plane of the sky,
as seen by the distant observer, is given by the
well known formula for superluminal motion (Rees 1966, 1967)
\begin{equation}
v_\mathrm{app}=\frac{v \sin\theta}{1\mp\beta\cos\theta}.
\end{equation}

After the blobs have traveled a physical
distance $l_\mathrm{jet}$ corresponding to a projected distance
$x_\mathrm{jet} = l_\mathrm{jet} \sin{\theta}$ along
the jet and counterjet they cease radiating.
The lifetime of these blobs is therefore given by $\tau=l_\mathrm{jet}/v$.
Due to time travel effects the apparent lifetime
$\tau_\mathrm{app}$
of the approaching
blob differs from the one of the receding blob,
which can be expressed as (e.g.~Ryle \& Longair 1967)
\begin{equation}
\tau_\mathrm{app}=\tau(1\mp\beta\cos\theta).
\end{equation}
The lifetime in the reference frame of a blob would introduce
an additional factor of $\gamma$ because of the transversal Doppler effect.

After a period of time $T$ a second pair of
blobs is emitted in our model and so forth.
If the AGN keeps producing blobs at such a constant rate,
the spatial separation between blobs will remain constant.
The apparent distance between two successive blobs
$\Delta x_\mathrm{app}$
is simply the product of their
apparent velocity and the period of the blob ejection cycle, viz.
\begin{equation}
\Delta x_\mathrm{app} = v_\mathrm{app} T =
\frac{\Delta x}{1\mp\beta\cos\theta}.
\end{equation}
Obviously, the number of blobs per unit length of jet scales
inversely proportional to this quantity. Since we are assuming the jet and
the counterjet to be of equal lengths, this quantity is also proportional to
the number of apparently active blobs at a given moment in time.

In the following sections the above-mentioned quantities are
required in order to compute the apparent
luminosity of the jet and the counterjet.
The total intrinsic jet luminosity is just the sum of the individual
blob luminosities of existing blobs.
In our model we try to approach the limit of
the radiation field produced by a continuous jet by increasing
the number of blobs per unit length of the jet
while keeping the assumed luminosity of the jet constant.

For optical thin conditions, which is a valid approximation in
the present case,
the scattered surface brightness at coordinates
$x$, $y$ on the plane of the sky and
at a time of observation $t$ is given by
the integral over the emissivity due to scattering
along the corresponding line of sight
\begin{equation}
I_{\nu}^\mathrm{sc}(x,y,t)=
\int_{z_\mathrm{min}}^{z_\mathrm{max}}
j_{\nu}^\mathrm{sc}(\textbf{r},t){\rm d}z,
\end{equation}
where $\textbf{r}$ is the radius vector of the point of scattering and
$z_\mathrm{min}$ and $z_\mathrm{max}$ are the boundaries of integration,
which as will be described in more detail further below
themselves depend upon the position on the plane of the sky
and the time of observation, i.e.~$x$, $y$ and $t$.

In the optical thin approximation the emissivity
due to scattering can be expressed as
\begin{equation}
j_{\nu}^\mathrm{sc}(\textbf{r},t)
=n_\mathrm{e}(\textbf{r})\sum_\mathrm{blobs}
\frac{P_{\nu}(\textbf{n},t')}
{|\textbf{r}-\textbf{r}_\mathrm{blob}(t')|^2}
\left(\frac{{\rm d}\sigma}{{\rm d}\Omega}\right),
\end{equation}
where $n_\mathrm{e}(\textbf{r})$ is the electron number density at
the position $\textbf{r}$,
$P_{\nu}(\textbf{n},t')$ the spectral power per unit solid angle emitted by
a blob at a ``retarded time'' $t'$ at a position
$\textbf{r}_\mathrm{blob}(t')$ into a direction defined by the
unit vector $\textbf{n}$ and $({\rm d}\sigma/{\rm d}\Omega)$ the differential
scattering cross section.
The unit vector pointing from the retarded
position of the blob towards the point of scattering is given by
\begin{equation}
\textbf{n}=\frac{\textbf{r}-\textbf{r}_\mathrm{blob}(t')}
{|\textbf{r}-\textbf{r}_\mathrm{blob}(t')|}.
\end{equation}

In order to evaluate the scattered surface brightness
at a time of observation $t$, the corresponding
retarded time $t'$ has to be computed.
The received photon must have traveled from its point of emission,
i.e. the position of the blob at a time $t'$,
via the point of scattering to the distant observer.
The relation between the time of observation $t$ and the retarded
time $t'$, therefore, is given by the following expression
\begin{equation}
c(t-t')=d-z+|\textbf{r}-\textbf{r}_\mathrm{blob}(t')|.
\end{equation}
Solving this implicit equation for $t'$ yields
\begin{equation}
t'=\gamma^2\frac{A-\sqrt{A^2-c^2B/\gamma^{2}}}{c^2},
\end{equation}
with
\begin{eqnarray}
A=c^2t-v_{x}(x-x_\mathrm{o})-v_{z}(z-z_\mathrm{o})-c(d-z)
\nonumber
\end{eqnarray}
and
\begin{eqnarray}
\nonumber
B=(c\,t-(d-z))^2-((x-x_\mathrm{o})^2+y^2+(z-z_\mathrm{o})^{2}).
\end{eqnarray}
Eq.~9 enables one to compute the position
of the blob at a time $t'$
as a function of the time of observation
$t$ and the point of scattering with
coordinates $x$, $y$ and $z$.

We assume a blob to emit synchrotron radiation
with a spectral index $\alpha$ (in the following
sections we will adopt a value of $\alpha=0.5$,
which is commonly observed in the spectra of radio jets).
Thus, the spectral luminosity is given by
\begin{equation}
L_{\nu}=L_{\nu}^\mathrm{o}\left(\frac{\nu_\mathrm{o}}{\nu}\right)^{\alpha}.
\end{equation}
Each blob radiates at an intrinsic rate of
$L_{\nu}^\mathrm{o}=4\pi P_{\nu}^\mathrm{o}$.
We assume the intrinsic luminosity of a blob to remain
constant over its lifetime.
The relation between the spectral luminosity per unit solid angle
in the observer's frame, i.e. the rest frame of the scattering electron
or equivalently the rest frame of the distant observer,
and in the rest frame of a blob is given by
(e.g.~Rybicki \& Lightman 1979)
\begin{equation}
P_{\nu}(\textbf{n})=\delta^{3+\alpha}P_{\nu}^\mathrm{o}=
\left(
\frac{1}{\gamma(1-\beta\cos\psi)}
\right)^{3+\alpha}
P_{\nu}^\mathrm{o},
\end{equation}
where $\delta$ is the Doppler or beaming factor.

The angle between the direction of motion of the blob
and the line connecting
the retarded position of the
blob, i.e.~the position of the blob at a time $t'$,
and the point of scattering $\psi$ can be computed as
follows. The cosine of this angle
is given by the scalar product of $\textbf{n}$
and the unit vector along the jet axis given by
$\textbf{e}_\mathrm{jet}=(\sin\theta,0,\cos\theta)$.
One thus obtains
\begin{equation}
\cos \psi=\frac{\sin{\theta}(x-x_\mathrm{blob})+
                \cos{\theta}(z-z_\mathrm{blob})}
                         {\sqrt{(x-x_\mathrm{blob})^2+
                                y^2+
                                (z-z_\mathrm{blob})^2}},
\end{equation}
where $x_\mathrm{blob}$ and $z_\mathrm{blob}$ are computed
according to Eq.~1.

Since we are considering photons produced by the synchrotron
process and since the scattering electrons have a temperature of
several keVs, it is appropriate to use
the differential cross section for Thomson scattering, i.e.
\begin{equation}
\left(\frac{{\rm d}\sigma}{{\rm d}\Omega}\right)=
\frac{3\sigma_\mathrm{T}}{16\pi}(1+\cos^2\phi),
\end{equation}
where $\sigma_\mathrm{T}$ is the Thomson cross section.

The angle between the line connecting the retarded blob position
and the line of sight can be computed analogously to the angle $\psi$.
The relation $\cos\phi=\textbf{e}_{z}\cdot\textbf{n}$,
where $\textbf{e}_{z}$ denotes the unit vector along the z-axis,
yields
\begin{equation}
\cos\phi=\frac{z-z_\mathrm{blob}}{\sqrt{(x-x_\mathrm{blob})^2+
y^2+(z-z_\mathrm{blob})^2}}.
\end{equation}

The density profile of the hot electron gas surrounding an AGN
located inside a central galaxy of a cluster can be derived
by means of the observed X-ray surface brightness profile
produced by this gas.
Since the computations in our model are generally performed
numerically, any density distribution
could be used in principle.
However, for the sake of simplicity
we are assuming for the time being a density distribution according
to the widely used beta profile
(Cavaliere \& Fusco-Fermiano 1976)
\begin{equation}
n_\mathrm{e}(r)=\frac{n_\mathrm{e}^\mathrm{o}}
{(1+(r/r_\mathrm{c})^2)^{\frac{3}{2}\beta_\mathrm{c}}},
\end{equation}
which is fully parametrised by the normalisation of the electron
number density
$n_\mathrm{e}^\mathrm{o}$, the core radius $r_\mathrm{c}$ and the
$\beta_\mathrm{c}$ parameter.
The application of our results to
the jet of M87 in a later section, however,
will use a recently determined
density profile for this object.

Finally the integration boundaries $z_\mathrm{min}$
and $z_\mathrm{max}$ of the integral occurring in Eq.~5
have to be computed. These quantities are functions of
the position on the plane of the sky $x$, $y$
and the time of observation $t$ and can be considered
as two surfaces evolving with time. These surfaces are defined
as the loci of scattering sites giving a fixed time delay
for photons traveling from their point of emission
to the distant observer, which divide space into
three distinct regions: one region that scatters photons
originating from the blob during its existence and two regions
that cannot scatter any photons since either the source has not
turned on yet or has already switched off.

For a stationary source of radiation the shape of these surfaces
is that of a paraboloid\footnote{It should be noted that the paraboloid is an
approximation to the more general case of an ellipsoid,
which is valid for the case of the source of radiation
and the scattering site being at large distances from the
observer.
The more general case has been applied
e.g.~in a study of the scattering of radiation produced by the
supermassive black hole at the centre of the Milky Way
by giant molecular clouds located in the Galactic molecular ring
(Cramphorn \& Sunyaev 2002).}
with the radiation source at the apex, which can be described
e.g.~by the following formula
(e.g.~Sunyaev \& Churazov 1998)
\begin{equation}
z=\frac{1}{2}\left(\frac{x^2+y^2}{ct}-c^2(t-d/c)^2\right).
\end{equation}

In the case of a moving radiation source $z_\mathrm{min}$
and $z_\mathrm{max}$ can be computed by setting
Eq.~9 equal to $t_\mathrm{min}'$ and $t_\mathrm{max}'$, respectively,
where $t_\mathrm{min}'$ is the time the blob started radiating and
$t_\mathrm{max}'$ the time the blob stopped radiating.
Solving this equation for $z_\mathrm{min}$ and $z_\mathrm{max}$,
respectively, it turns out that $z_\mathrm{min}$ also can be
described by Eq.~16. This, however, is actually not surprising,
because our model was set up in a way that a blob is created
in the nucleus of the AGN corresponding to our origin of
coordinates.

Concerning $z_\mathrm{max}$ the result is just a generalisation of
Eq.~16 describing the loci of equal travel times from a stationary
source located at the origin to the observer, i.e.
\begin{eqnarray}
z_\mathrm{max}=
\frac{1}{2}
\Big(
\frac{(x-x_\mathrm{blob}(t_\mathrm{max}'))^2+y^2+
z_\mathrm{blob}(t_\mathrm{max}')^2}{ct}-
\nonumber
\\
c^2(t_\mathrm{max}'-(t-d/c))^2
\Big).
\end{eqnarray}
This equation describes a similar paraboloid as Eq.~16 except
that its apex corresponds to the point in space where the blob
ceases radiating.

With the above equations at hand the remaining problem 
turns out to be a function of the variables 
$x$, $y$, and $t$, the free parameters being 
the intrinsic spectral luminosity of the blob or jet
$L_{\nu}^\mathrm{o}$ , the bulk Lorentz factor $\gamma$, the inclination
angle $\theta$, the parameters of the electron density 
distribution, i.e. $n_\mathrm{e}^\mathrm{o}$, 
$\beta_\mathrm{c}$ and $r_\mathrm{c}$, 
and the length of the jet $l_\mathrm{jet}$, 
which together
with a given Lorentz factor corresponds 
to a given lifetime of a blob.

\section{Results}
Using Eq.~5 one is now in a position to compute the surface brightness
of the diffuse scattered emission for a given scenario
as a function of the position on the plane of the sky 
and the time of observation.   
In the following sections we will use the brightness temperature
more commonly used in radio astronomy to express our results, i.e.
\begin{equation}
T_\mathrm{b}^\mathrm{sc}=\frac{\lambda^{2}}
{2k_\mathrm{B}}\,I_{\nu}^\mathrm{sc},
\end{equation}
where $k_\mathrm{B}$ is Boltzmann's constant. A wavelength of observation 
of 6 cm is assumed corresponding to a frequency of about 5 GHz.

\subsection{Single ejection}
A first scenario to apply our model to is the single
ejection of a pair of blobs.
In order to compare the results of the present section
with the results of later sections 
we have chosen the following set of parameters as a 
reference model which may be representative
of an ``average'' jet originating in the nucleus of
the central galaxy of an ``average'' 
cluster of galaxies:
$\gamma=5$, $\theta=30^{\circ}$, 
$L_\mathrm{\nu}^{\mathrm{o}}=10^{32}\,{\rm
erg}\,{\rm s}^{-1}\,{\rm Hz}^{-1}$, 
$l_\mathrm{jet}=2\,{\rm kpc}$,
$n_\mathrm{e}^{\mathrm{o}}=10^{-1}\,{\rm cm}^{-3}$, $\beta_{\mathrm{c}}=2/3$ 
and $r_\mathrm{c}=10\,{\rm kpc}$. 
Furthermore, we place the source at a distance of 16 Mpc 
from the observer.
The corresponding observed flux is about 
$8.5\times10^{2}$ Jy. 
For the chosen set of parameters the
blobs have a ``lifetime'' 
of about $6.800$ years
corresponding to an apparent lifetime of 
the approaching blob of about 
$1.000$ and 
of the receding blob of about $12.000$ years, respectively. 
Using Eq.~2 the apparent velocities can be computed 
and turn out to be 
3.23c (about 1.000 parsec per 1.000 years) 
for the approaching blob and 
0.27c (about 80 parsec per 1.000 years) 
for the receding blob,
respectively.
As can be taken from Eqs.~5 and 6 our results easily can be 
converted to a different density normalisation
or a different intrinsic luminosity, since
the scattered surface brightness just scales linearly with 
these quantities.

\begin{figure*}
\centering
\includegraphics{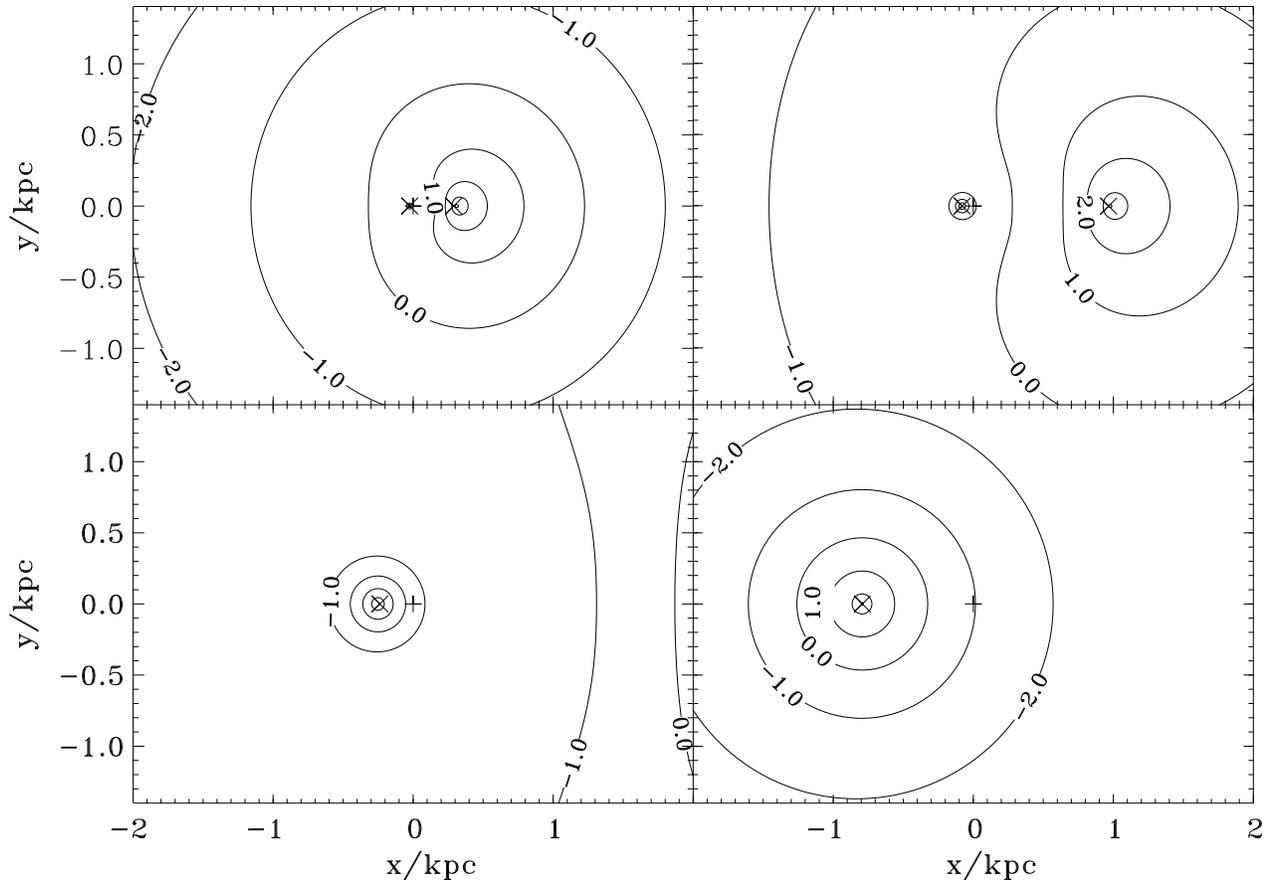}
 \caption{Contour plots of the brightness temperature of the scattered
          radiation three hundred (upper left), one thousand (upper
          right), three thousand (lower left) and 
          ten thousand (lower right) years after the ejection of a
          pair of blobs with an inclination angle of
          $\theta=30^{\circ}$ and a Lorentz factor of $\gamma=5$.
          The blob moving to the right is approaching
          the observer. Beyond a projected distance of 1 kpc from the nucleus
          the blobs cease to radiate.
          The positions of the nucleus and the approaching and the receding blob
          are marked by a plus sign and by crosses, respectively. 
          Logarithmic contour levels are shown down to $\log T_\mathrm{b}=-2$.}
\label{fig2}
\end{figure*}

\subsubsection{Brightness Temperature}
In Fig.~2 we are plotting
contours of equal brightness temperature of
the scattered radiation in the plane of the sky 
three hundred, one thousand, three thousand and ten thousand years
after the ejection of a pair of blobs 
out of the nucleus of an AGN
for the above chosen set of parameters. 
The two blobs travel a projected distance of 1 kpc along the 
positive and negative x-axis, respectively, and then stop to
radiate, wherein the blob moving to the right is approaching 
the observer.

In the first snapshot three hundred years after 
the ejection of the pair of blobs the scattered radiation due 
to the approaching blob is already rather extended,
whereas the scattered radiation
produced by the receding blob is barely visible 
at the centre. 
The scattered surface brightness stemming from the approaching
blob has already reached a maximum value of more than
$10^4$ Kelvin and the position of this maximum
coincides with the apparent position
of the approaching blob.
This maximum is about 300 pc
distant from the nucleus. 
Although the distribution of the scattered
surface brightness due to the approaching blob is symmetric with
respect to the x-axis, such a symmetry is not given with respect
to the direction perpendicular thereto. Rather some contour lines  
of the scattered radiation produced by the approaching blob adopt 
a more ``kidney-like'' shape,
which becomes even more pronounced in the following snapshot
seven hundred years later. We have observed this shape in
our numerical computations
in particular for the cases of small inclination angles
and large Lorentz factors. However, the contour lines at further
distances from the nucleus exhibit a spherically symmetric shape.

The second snapshot one thousand years after the ejection 
of the pair of blobs
shows the morphology of the brightness temperature of the scattered 
emission at a time just before the approaching blob 
switches off, i.e.~reaches
a projected distance of 1 kpc from the nucleus. 
Compared with the previous snapshot
the pattern of scattered radiation produced by the approaching blob
has spread out further, while the maximum brightness temperature
has decreased somewhat.
Close to the nucleus
the scattered radiation originating from the receding blob  
is just starting to
become more extended, but already has reached a maximum brightness
temperature of almost $10^4$ Kelvin. 
As can be taken from this and especially
the two following snapshots the scattering pattern produced by
the receding blob exhibits a more spherically symmetric shape than
the scattering pattern produced by the approaching blob.

Two thousand years later, i.e.~three thousand years after
the ejection of the pair of blobs, the maximum of scattered
radiation due to the approaching blob 
has already moved beyond 
the scales of Fig.~2, i.e.~to distances more than 2 kpc 
away from the nucleus. Only a single contour line corresponding to a
brightness temperature of 1 Kelvin is still visible
on the right side of the plot. 
As the approaching blob already has switched off,
this ``afterglow''
of scattered radiation keeps on moving away
from the nucleus with the same apparent velocity as the
approaching blob during its existence. 
The maximum of this afterglow actually 
is about 3 kpc distant from the
nucleus and will keep on moving to even larger distances,
as time progresses further, while at the same time 
decreasing in brightness until completely fading away.
The scattered radiation of the receding blob,
on the other hand, still
can be seen centered at a distance of about 250 pc 
to the left of the nucleus.

Finally, after ten thousand years the scattered 
radiation produced by the approaching blob fully 
has disappeared
(the maximum would be located at a distance of almost 10 kpc 
from the nucleus now) 
and only the pattern of scattered radiation produced by the receding 
blob can be observed. In the meantime,
this pattern of scattered radiation due to the
receding blob has become more extended, while traveling out 
to a distance of about 800 pc.
It should be noted that as in the case of the approaching blob, 
at still later times one can observe an ``afterglow'' of scattered radiation
produced by the receding blob even after this blob has stopped radiating.
Also this pattern of scattered radiation 
on the side of the receding blob 
keeps on moving with the same apparent
velocity as the receding blob itself.

\begin{figure}
\centering
\includegraphics[width=\columnwidth]{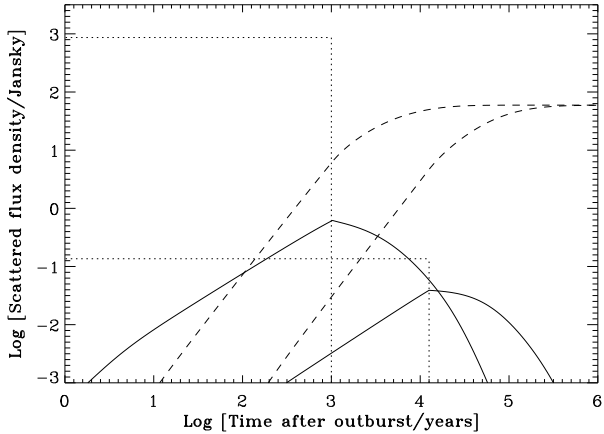}
 \caption{Temporal evolution of the scattered
          flux densities (solid lines), the apparent flux densities 
          (dotted lines) and the fluences of scattered flux (dashed lines)
          for the approaching and the receding blob shown in Figure 2.
          As long as the approaching blob keeps on radiating the
          ratio of its apparent flux density to its scattered flux
          density is larger than three orders of magnitude. 
          After the blobs switch off (reflected by the kinks in the
          lightcurves of the scattered flux and the drop off of
          the apparent direct flux)
          an ``afterglow'' of scattered radiation remains.
          The fluence of the scattered flux 
          for both the approaching as well as for the
          receding blob reaches a maximum value of about 
          $2.4\times10^{-16}\,{\rm erg}\,{\rm cm}^{-2}\,{\rm Hz}^{-1}$.}
\label{fig3}
\end{figure}

As already mentioned and as can be taken from Fig.~2,
the respective maxima
of the scattered surface brightnesses coincide with the respective apparent
blob positions and, thus, both move with the same apparent velocity.
We have verified that this holds true 
even for very large Lorentz factors, 
i.e.~that the highest brightness
temperature is found at the apparent position of a blob.
For large Lorentz factors the scattered emission 
tends to become more ``concentrated'' around a blob. 

The above results differ somewhat from the findings of GSC87,
who showed that for the approximation of two
oppositely directed radiation cones a burst 
(equivalent to our blob lifetime) produces 
two pronounced maxima in the brightness distribution
(see Fig.~2 of GSC87), wherein the maximum on the
side of the cone pointing in the direction
of the observer is fainter than the one on the side
of the radiation cone pointing away from the observer.
Such a behaviour was not observed in our computations.
Furthermore, it was shown in GSC87, that using the adopted
approximation the velocities of the fainter maximum  
and the brighter maximum can be given by a formula
solely depending on the inclination angle. Obviously, this
approximation does not allow to derive any dependence
on the Lorentz factor. As we have seen in our computations,
however, the velocity of the scattering pattern produced by a blob
is given by Eq.~2.

\subsubsection{Flux and Fluence}
With respect to our model a physically somewhat more 
``global'' quantity than
the scattered surface brightness
is the total scattered flux originating from a blob.
In Fig.~3 we are plotting the scattered 
flux densities, the apparent flux densities and the fluences 
produced by each blob of the  
pair of blobs shown in Fig.~2
as a function of time up to $10^{6}$ years after the ejection.
The fluxes are measured in units of Janskys, where 
$1\,{\rm Jy}$ corresponds to a flux of 
$10^{-23}\,{\rm erg}\,{\rm cm}^{-2}\,{\rm s}^{-1}\,{\rm Hz}^{-1}$.

The general behaviour of the two blobs in time as exhibited in
the contour plots shown in Fig.~2 and as described in the 
previous section is also reflected by the scattered fluxes
produced by these blobs.
Initially, the scattered flux due to the approaching blob
dominates and increases linearly in time. About one thousand years
after the ejection corresponding to the time when the approaching
blob switches off, its scattered flux peaks at a maximum value
of about 600 mJy. After the switch off an ``afterglow'' of 
scattered radiation remains for some time at a similar
level of flux.

The lightcurve 
corresponding to the scattered flux due to the receding blob 
is similar in shape, however, peaks at a smaller maximum value
and is shifted to later times. The scattered flux produced by 
the receding blob reaches its maximum value of about 40 mJy
about 12.000 years after the ejection and remains at a similar
flux level for almost $10^{5}$ years after the switch off
of the receding blob. 

In order to estimate a characteristic size of the distribution
of the scattered radiation produced by these blobs,
i.e~the size of the region where the bulk of the scattered flux originates
from, we have computed
the flux coming from a circle of radius r as a function
of this radius, wherein the circle is centered
on the position
of the blob or equivalently on the position of the maximum
of the surface brightness distribution of the scattered 
radiation. Our results show that the size of this characteristic region 
grows with time, remains, however, smaller than about 1 kpc 
for about $3\times10^{3}$ years for the approaching blob and for about
$3\times10^{4}$ years for the receding blob.

\begin{figure}[t]
  \centerline{\psfig{figure=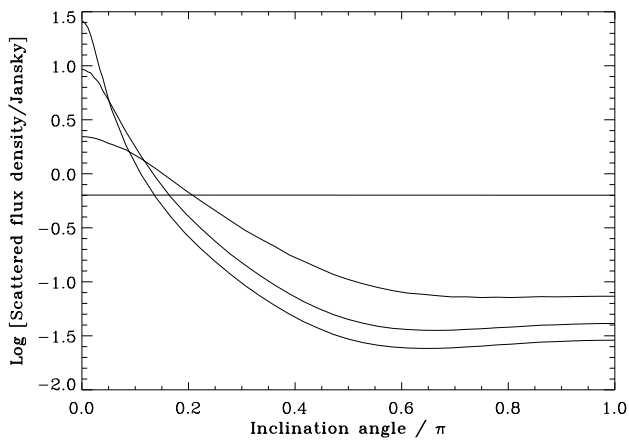,width=\columnwidth}}
 \caption{Maximum of the scattered flux density produced by a single
          blob as a function
          of the inclination angle for four different Lorentz factors, 
          viz.~1 (stationary source), 2, 5 and 10 
          (otherwise the same parameters have been used as before).
          For small inclination angles the blob moving with a Lorentz
          factor of 10 is producing the largest scattered flux.
          Note that for very large inclination angles the maximum of
          the scattered flux slightly starts to increase again. This
          effect is caused by the angular dependence of the differential cross
          section for Thomson scattering which ``amplifies'' forward
          as well as backward scattering compared to a scattering at
          a right angle.}
\end{figure}

\begin{figure}[t]
  \centerline{\psfig{figure=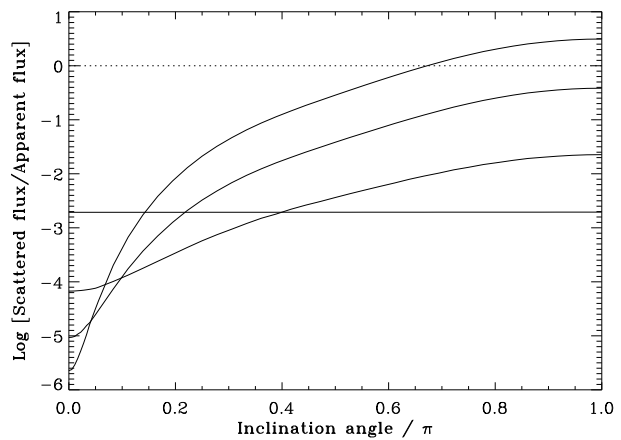,width=\columnwidth}}
 \caption{Ratio of the maximum of the scattered flux density to the 
          apparent flux for a single blob as a function of 
          the inclination angle of
          the direction of motion of the blob with respect to the 
          line of sight for four different Lorentz factors, 
          viz.~1, 2, 5 and 10. The ratio for the stationary source
          is an estimate of the optical depth for Thomson scattering
          of the surrounding gas, which in this case
          is about $1.85\times10^{-3}$. 
          Given the assumed electron density and the
          corresponding Thomson
          depth,
          the ratio of the scattered flux to the apparent flux
          reaches unity
          at an inclination angle
          of about $2/3 \pi$ for a Lorentz factor of 10. 
          For even larger inclination angles the
          scattered flux becomes larger than the apparent flux.}
\end{figure}

In order to set the values for the scattered fluxes into context,
we have plotted the apparent ``direct'' fluxes received from
the approaching blob as well as from the receding blob for a comparison. 
Obviously these direct fluxes vanish once the blobs stop to radiate,
as reflected by the drop off in flux after about 1.000 years and
12.000 years, respectively.
While the approaching blob is active, i.e.~radiating,
the ratio of its apparent direct flux to its scattered flux is
about 1.400. For the receding blob the ratio of the
apparent direct flux to the scattered flux turns out to be
a factor of about 3. 

As we have seen the approaching and the receding blob
exhibit a different behaviour, 
when studying the surface brightness profiles
produced by these blobs (Fig.~2) 
or their scattered flux (Fig.~3). 
The brightness temperature of the scattered radiation
produced by the approaching blob reaches higher values 
and is more extended than in the case of the receding blob.
Thus,
the corresponding scattered flux produced by the approaching blob
for a given time
is larger than the scattered flux 
of radiation originating from the receding blob,
as depicted in Fig.~3, provided that the blobs
are still active. 
However, as also exhibited in Figs.~2 and 3 the apparent 
lifetime of the receding blob is correspondingly longer
than the apparent lifetime of the approaching blob.
It is, thus, interesting to look at the total time
integrated fluxes produced by these blobs. 

\begin{figure}[t]
  \centerline{\psfig{figure=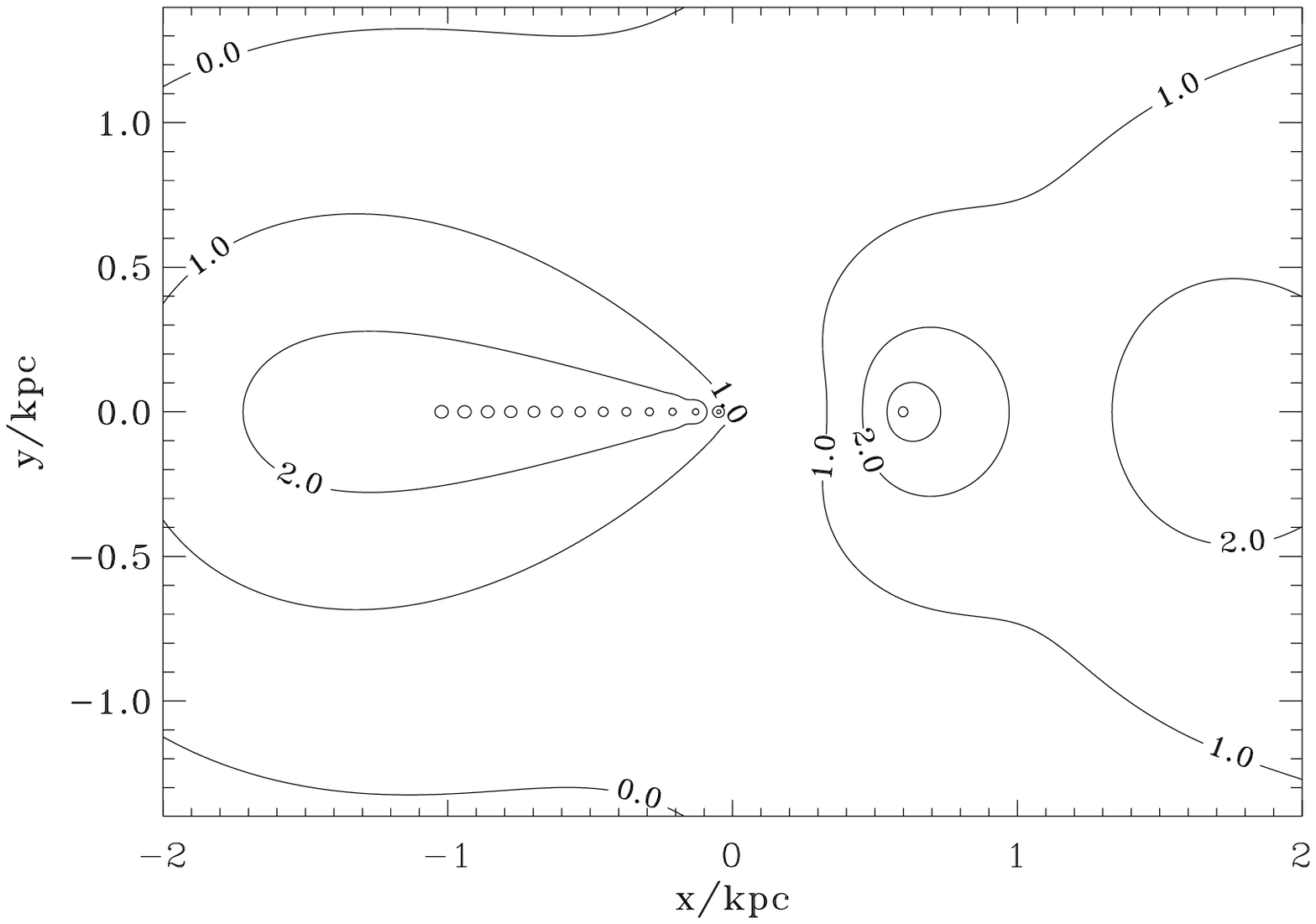,width=\columnwidth}}
 \caption{Contour plot of the brightness temperature of the scattered
          radiation about twenty thousand years after the onset of the jet
          activity, i.e.~the ejection of the first 
          pair of blobs. About every 1.000 years a further pair of
          blobs is being produced. The corresponding apparent distance between individual
          blobs is about 1 kpc for the approaching blobs on the right
          side of the nucleus and about 80 pc for the receding blobs
          on the left side of the nucleus.}
\end{figure}

To this end, Fig.~3 also shows the respective
fluences, i.e.~the time integrated fluxes,
of the approaching and the receding blob. As can be taken
from this figure, these fluences approach the same values
as expected.

\begin{figure*}
\centering
\includegraphics{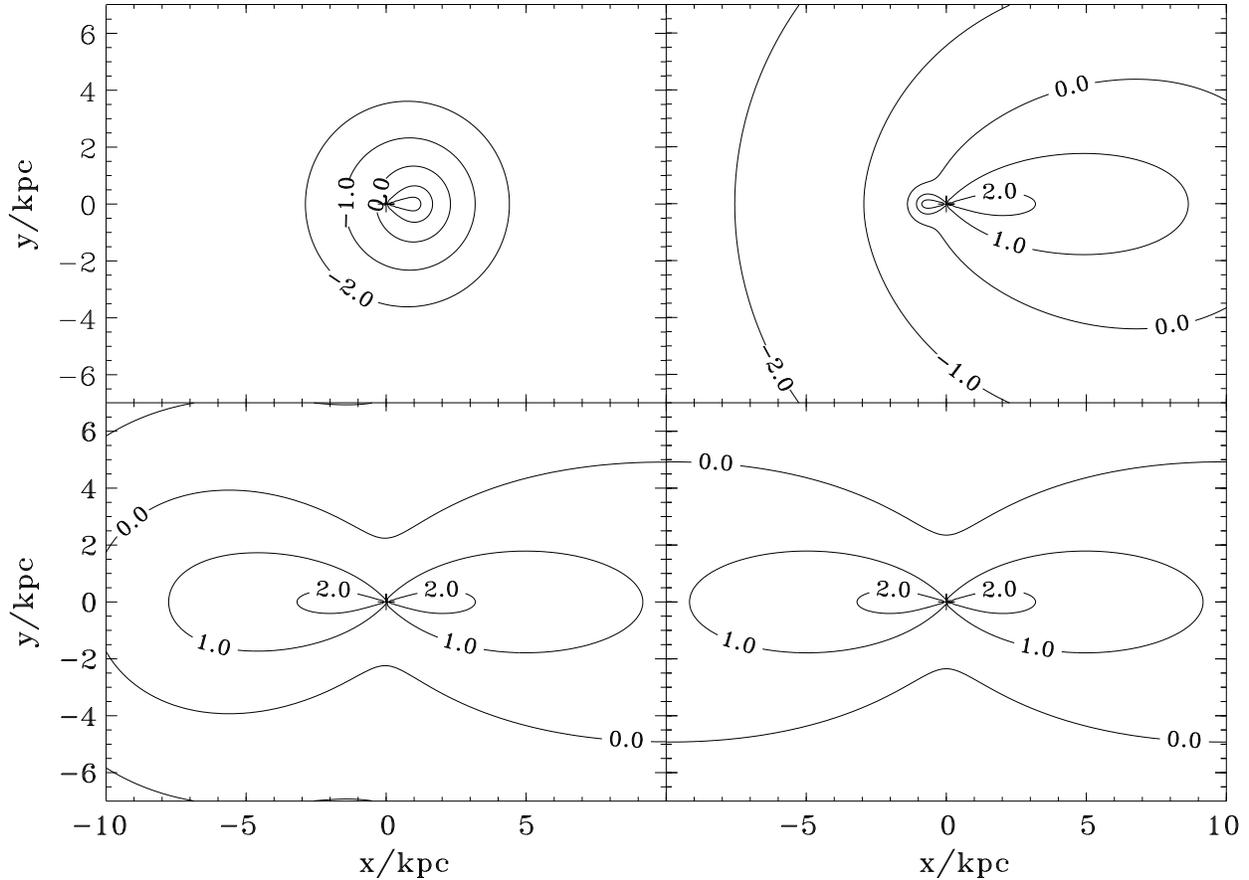}
 \caption{Contour plots of the brightness temperature of the scattered
          radiation $10^{3}$ (upper left), $10^{4}$ (upper right), 
          $10^{5}$ (lower left) and 
          $10^{6}$ (lower right) years after the onset of the jet
          activity, i.e.~the ejection of the first 
          pair of blobs. About every 170 years a further pair of blobs
          has been ejected.      
          The position of the nucleus is marked by a plus sign.
          Logarithmic contour levels are shown down to $\log T_\mathrm{b}=-2$.}
\end{figure*}

We now turn to the dependency of the maximum scattered flux produced
by a single blob on the inclination angle and the Lorentz factor.
Fig.~4 shows this maximum scattered flux for the case of a stationary
blob (which obviously shows no dependence on the inclination angle)
and three blobs moving with different Lorentz factors. 
For small inclination angles
the blob with the largest Lorentz factor, i.e.~$\gamma=10$, 
produces the largest scattered flux,
which decreases for larger
inclination angles.
It is interesting to note, however, that beyond a certain inclination angle 
the scattered flux slightly starts to increase again. We have verified that
this behaviour is caused by the specific
angular dependence of the Thomson scattering cross section
which enhances forward and backward scattering compared
with scatterings at different angles. 

\begin{figure*}
\centering
\includegraphics{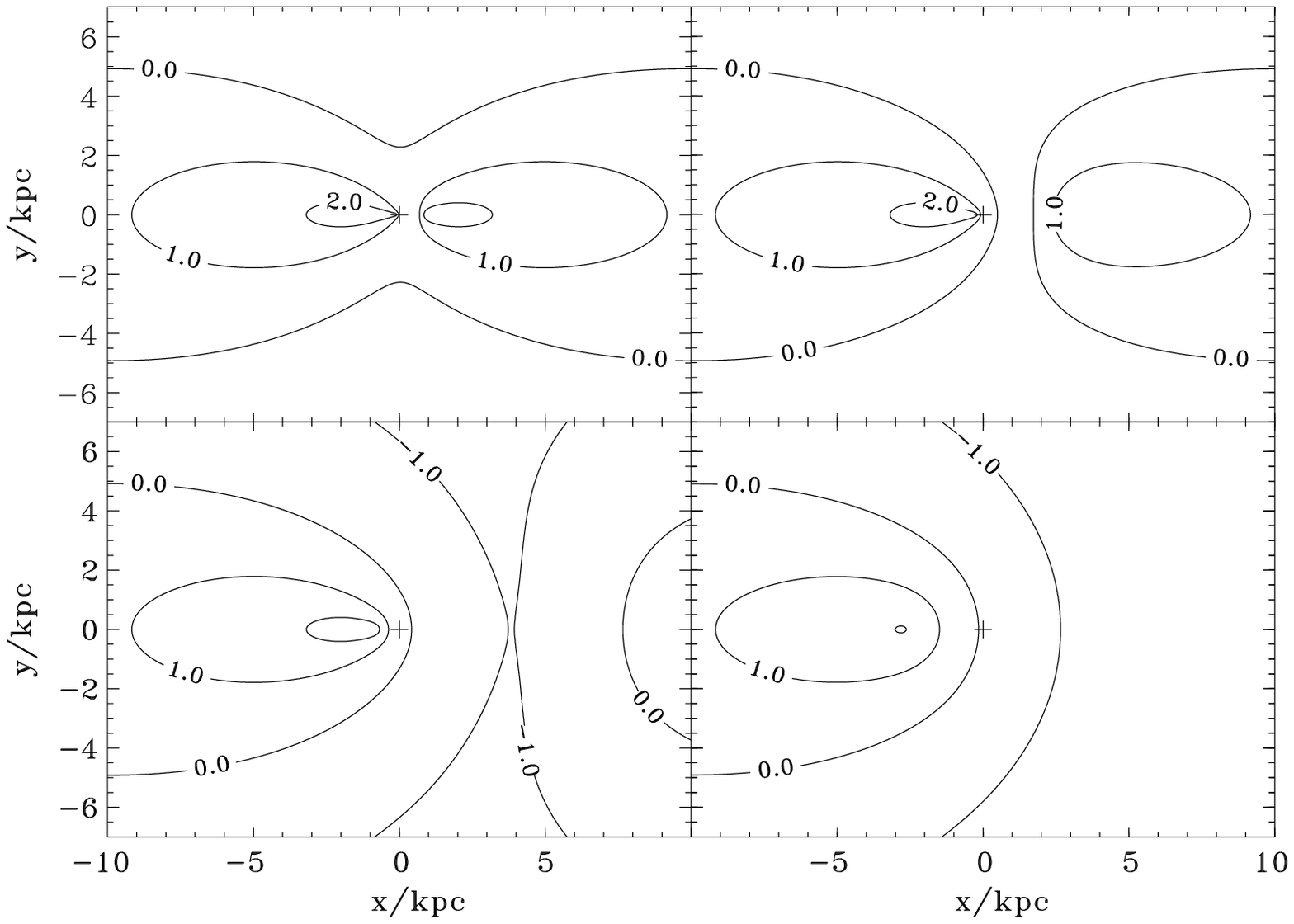}
 \caption{Contour plots of the brightness temperature of the scattered
          radiation one thousand (upper left), three thousand (upper
          right), ten thousand (lower left) and 
          thirty thousand (lower right) years after the demise of the jet
          activity, i.e.~the ejection of the last 
          pair of blobs, which lasted for $10^{6}$ years
          (corresponding to the snapshot shown in the lower right of Fig.~7).         
          The position of the nucleus is marked by a plus sign.
          Logarithmic contour levels are shown down to $\log T_\mathrm{b}=-2$.}
\end{figure*}

In order to be able to address the question, whether the scattered
radiation discussed in this section might be observable at all, 
we have plotted in Fig.~5 the ratio 
of the total scattered flux produced by a blob to its
apparent direct flux as a function of the inclination angle
for four different Lorentz factors, 
viz.~1 (stationary source), 2, 5 and 10. 
Since for a stationary point source the scattered flux is given by
$F_\mathrm{sc}=\tau_\mathrm{T} F_\mathrm{o}$
(Sunyaev 1982), this ratio should correspond to the
optical depth for Thomson scattering. For  
the chosen electron density distribution the
Thomson depth turns out to be about $1.85\times10^{-3}$.

Concerning the case of a moving blob,
it can be taken from Fig.~5 that 
for small inclination angles the scattered flux is several
orders of magnitude smaller than the direct flux,
wherein this ratio decreases for larger Lorentz factors. 
As one goes to larger inclination angles, however,
the ratio of the scattered flux
to the direct flux increases substantially. 
For a Lorentz factor of 10
this ratio reaches unity at an inclination angle of about
$120^{\circ}$. For even larger inclination angles
the flux of scattered radiation
becomes larger than the flux of direct radiation given
a Lorentz factor of 10. 
For Lorentz factors of 5 and below the ratio does not
reach unity for the chosen set of parameters. However,
since the scattered radiation is simply proportional to the electron 
density and the direct emission does not depend thereon, the ratio 
of these two quantities scales linearly with the assumed electron
density.

\subsection{Multiple ejections}

In this section we will study a multiple blob scenario, 
i.e.~instead of creating only a single pair of blobs  
a second pair of blobs is ejected after a certain period 
of time 
and so forth. If the ejection period is chosen to be short enough, 
the ambient scattering medium will ``sense'' no difference
to a quasi-continuous ejection of radiating matter
moving at relativistic velocities.
As will become clear from the results presented in this section
and as can be taken from Eqs.~5 and 6, the problem we are studying
is linear in the contribution of each single blob.
Thus, the scattering pattern produced by several blobs essentially 
is just the sum of several spatially and temporally shifted 
scattering patterns produced by a single blob
as illustrated in Fig.~2.

\subsubsection{Brightness Temperature}
Fig.~6 shows a contour plot of the brightness 
temperature of the scattered radiation 
about 20.000 years after the ejection 
of a first pair of blobs for the multiple blob scenario. 
A pair of blobs is being 
ejected every 1.000 years corresponding 
to an intrinsic spatial blob separation of about 300 pc. 
Given a Lorentz factor of 5 and and an inclination
angle of $30^{\circ}$, the apparent distances between
two subsequent
blobs can be computed according to Eq.~4 
yielding a distance of about 1 kpc between approaching blobs
and about 80 pc
between receding blobs, respectively. These different apparent
distances between blobs for the jet and the counterjet side are 
readily apparent in Fig.~6. 
As before, we assume the blobs to switch off after 
having traveled a projected distance of 1 kpc.
As, furthermore, can be taken from Fig.~6, 
the pattern of scattered radiation on the side of the
receding blobs looks much ``smoother'' than the one on the jet side.
However, as will become clear from the following figures, 
by reducing
the blob separation further, also the scattering pattern
on the side of the approaching blobs will adopt a more or less
smooth shape. 

In the following it is our aim to study the behaviour exhibited in Fig.~6
for a larger number of blobs, for longer times and on larger scales.
To this end, we are plotting in Fig.~7
the brightness temperature of the scattered radiation
as a function of 
position on the plane of the sky at $10^{3}$, $10^{4}$, 
$10^{5}$ and $10^{6}$ years 
after the onset of the assumed jet activity.
Again, the blobs are ejected
with an inclination angle of $30^{\circ}$ 
with respect to the line of sight, move with a velocity 
corresponding to a Lorentz factor
of 5 and stop radiating beyond a projected distance of 1 kpc. 
An intrinsic spatial separation of the blobs of 50 pc 
has been chosen corresponding to an ejection of a pair of blobs about 
every 170 years.
Save to the larger spatial dimensions of the chosen field of view,
the contour plots shown in this figure are analogous 
to the contour plots shown in Figs.~2 and 6.

Over a time span of $10^{6}$ years we are observing the gradual
build up of a stationary scattering pattern initially on the side
of the jet and thereafter on the side of the counterjet.
In particular, it can be taken from the first snapshot
in Fig.~7 that
one thousand years after the onset of the jet activity the
scattered radiation pattern begins to develop on the
side of the approaching blobs, i.e.~the side of the jet. 
Already at these early times
the high brightness temperatures 
on the side of the approaching blobs 
adopt an elongated jet-like shape
close to the nucleus. 
The contours of lower brightness temperatures at larger distances 
from the nucleus exhibit a more symmetric shape. 
On this scale one cannot discern any contribution 
to the scattered radiation produced by
the blobs on the side of the counterjet yet. 

However, 9.000 years later scattered radiation produced by the 
receding blobs can be observed up to a distance of 1 kpc
from the nucleus exhibiting a jet-like shape similar to the
one produced on the side of the jet in the earlier snapshot.
The scattering pattern on the jet side,
on the other hand, 
has in the meantime almost fully developed on the scales shown
in this figure, 
especially in the vicinity
of the nucleus. Further generations of approaching blobs
will maintain this ``steady state''. 

One hundred thousand years after the onset of the jet activity, 
the picture is nearly symmetric with respect to the y-axis, i.e.~the
scattering patterns on the jet and the counterjet side
appear almost identical. Only for distances of about 10 kpc
to the left of the nucleus
the scattered radiation on the side of the counterjet is still
a bit fainter in comparison with the corresponding regions on the jet side.
In other words, the scattering pattern on the side of the counterjet
has not ``caught up'' completely with the scattering pattern
on the jet side yet, which as mentioned before
by this time already has reached
a stationary shape. 

\begin{figure}[t]
  \centerline{\psfig{figure=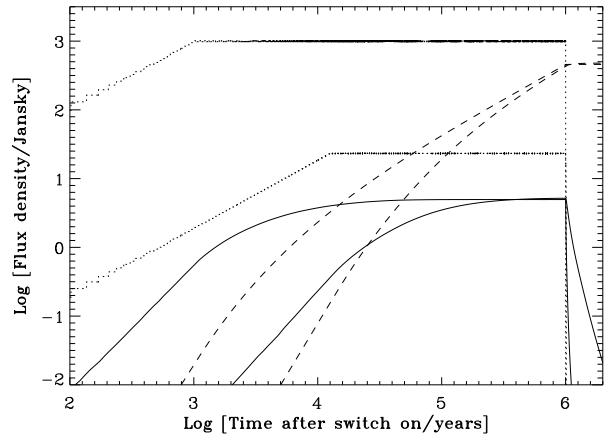,width=\columnwidth}}
 \caption{Temporal evolution of the scattered
          flux densities (solid lines), the apparent flux densities 
          (dotted lines) and the fluences of scattered flux (dashed lines)
          for the jet and the counterjet shown in Figs.~7 and 8.
          After a time of about $3\times 10^{5}$ years the counterjet produces
          the same amount of scattered flux as the jet as long
          as the source is active.
          The fluence of the scattered flux 
          for both the jet as well as for the counterjet reaches 
          a maximum value of about 
          $4.6\times10^{-13}\,{\rm erg}\,{\rm cm}^{-2}\,{\rm Hz}^{-1}$.}
\end{figure}

In the last snapshot of Fig.~7 illustrating the situation
$10^{6}$ years after the onset of
the jet activity
the whole pattern of scattered radiation
has reached a symmetric, stationary shape. 
Differently put, if we would keep producing blobs at the 
adopted rate, the morphology
exhibited by the
scattered radiation obviously would not change.
This ``steady-state'' scattering pattern
is reminiscent of the plot presented in GSC87 (see their Fig.~1),
which was 
obtained by a point source radiating into two oppositely
directed cones. Obviously, on the scales shown in Fig.~7 such
an approximation produces similar results, 
since the projected length of the 
jet and the counterjet of 1 kpc
is small compared to the dimensions
of the whole system illustrated in this figure.
Essentially, one can think of the final 
snapshot shown in Fig.~7  
being made up of
several ``overlapping'', more or less spherically symmetric patterns
produced by a single blob as shown in Fig.~2. 

Now that we have been producing blobs for $10^{6}$ years and
the scattered surface brightness exhibits a symmetric shape,
we want to study what happens when the jet switches off, 
i.e.~when no more blobs are being ejected. 
Fig.~8 shows the effect of such a switch off by means of 
contour plots of the brightness temperature
of the scattered radiation one thousand, three thousand, ten thousand
and thirty thousand years after the demise of the jet activity.

\begin{figure}[t]
  \centerline{\psfig{figure=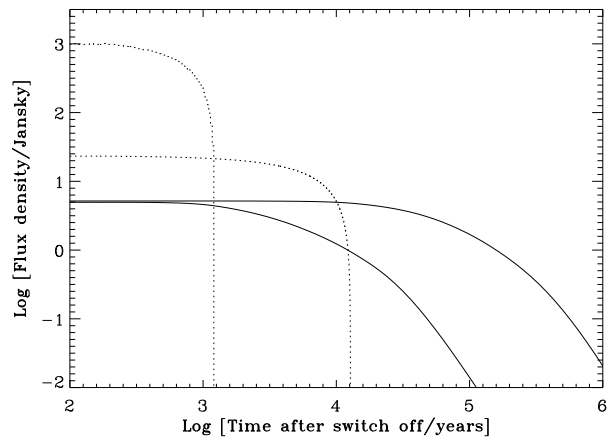,width=\columnwidth}}
 \caption{Temporal evolution of the scattered
          flux densities (solid lines) and the apparent flux densities 
          (dotted lines) for the jet and the counterjet 
          shown in Fig.~8 after
          the demise of the jet activity, which lasted for $10^{6}$ years.}
\end{figure}

Initially, the brightness temperature in the vicinity 
of the nucleus drops on the side of the jet. 
As time progresses, this region of reduced brightness moves
out to larger distances from the nucleus.
After about ten thousand years 
a similar behaviour can be observed
on the side of the counterjet.
By the time of thirty thousand years after the demise of the jet 
activity the
scattered emission on the side of the jet has vanished completely.
However, regions on the side of the counterjet being more distant from 
the nucleus still exhibit some brightness, wherein the maximum
brightness temperature is located at about 3 kpc to the left of the
nucleus. At even later times the maximum of this afterglow of
scattered radiation will have traveled to greater distances
from the nucleus and will have decreased further in brightness.  

\subsubsection{Flux and Fluence}
As in the section discussing the 
scenario with a single pair 
of blobs, we now turn to the
scattered fluxes in order to study how the general behaviour
exhibited in Figs.~7 and 8 is reflected by these quantities. 

\begin{figure*}
\centering
\includegraphics{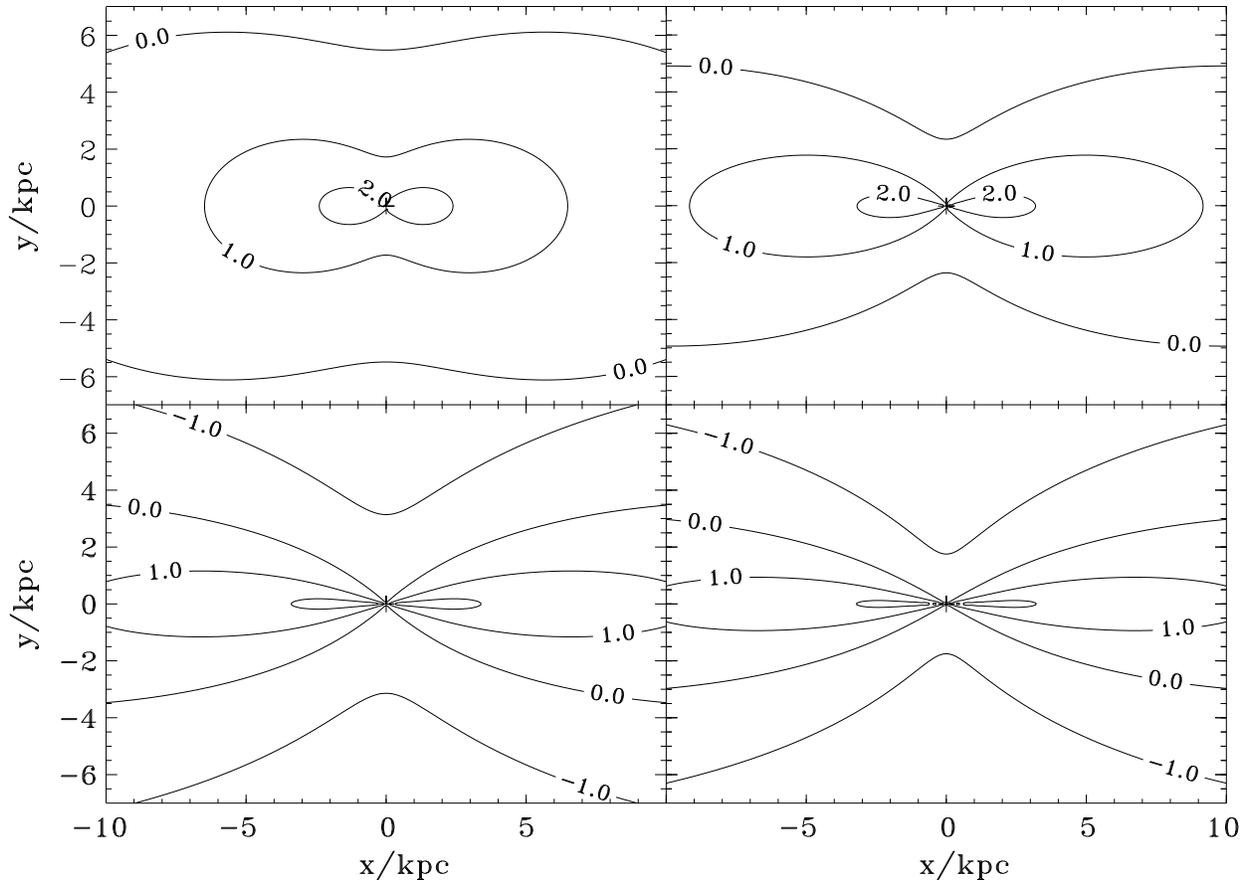}
 \caption{Contour plots of the brightness temperature of the
          scattered radiation for the stationary, symmetric case
          for an inclination angle of $15^{\circ}$ (upper left), 
          $30^{\circ}$ (upper right),
          $60^{\circ}$ (lower left) and 
          $90^{\circ}$ (lower right). The time required to 
          reach such a stationary state is roughly $4\times 10^{5}$, 
          $2.5\times 10^{5}$, $10^{5}$ and $5\times 10^{4}$
          years, respectively. The position of the nucleus is marked
          by a plus sign.}
\end{figure*}

Fig.~9 being similar to Fig.~3 shows the scattered flux,
the direct flux and the fluence produced by the jet and the
counterjet as a 
function of time after the onset of the jet activity
in our multiple blob scenario. Given the assumed 
set of parameters,
the scattered flux due to the counterjet reaches the same
level as the scattered flux due to the jet after 
about three hundred thousand years
corresponding to the stationary scattering pattern
illustrated in the snapshot on the lower right of Fig.~7.
However, as in the case of a single pair of blobs
initially the flux due to the
approaching blobs dominates over
the flux of scattered radiation
produced by the receding blobs.
It should be noted that 
the lightcurve shown in Fig.~9 is essentially just the sum 
of the lightcurves for a single blob
as shown in Fig.~3 shifted by a constant time difference.

The behaviour of the scattered fluxes after the switch off is 
difficult to assess from Fig.~9. We, therefore, have plotted
the scattered flux and the apparent flux after the switch-off in Fig.~10
using a different time axis.
As in the case of a single pair of blobs, one can observe an
afterglow of diffuse scattered radiation after the apparent fluxes
have died away that lasts substantially longer on
the side of the counterjet. Note that in this case the
switch-off of the jet is not a simple step function,
but rather the sum of multiple step functions shifted
in time. 

We now want to study the dependence of the quantities
discussed above
on the inclination angle and the Lorentz factor.
To this end Fig.~11 shows stationary
scattering patterns for four different inclination angles,
viz.~$15^{\circ}$, $30^{\circ}$, $60^{\circ}$ and $90^{\circ}$.
The other parameters have been chosen as in the previous
examples.
The scattering patterns in question
which exhibit a symmetry with respect to the 
x- and the y-axis are reached 
after a quasi-continuous ejection 
of blobs for roughly 
$4\times 10^{5}$, 
$2.5\times 10^{5}$, $10^{5}$ and $5\times 10^{4}$
years, respectively. 
These values were estimated by computing the time, when
the flux inside the region shown in the snapshots
of Fig.~11 reaches and remains at the same constant value on the
jet side as well as on the side of the counterjet.
For this reason in the case of $\theta=30^{\circ}$
the time to reach the steady state 
is a bit shorter than the time 
deduced from Fig.~9, because in that figure the total 
scattered flux is shown, to which regions beyond the scales 
of the snapshots given in Fig.~11 contribute.

The time needed to reach the stationary state obviously
increases with decreasing inclination angle.
The reason for this behaviour is as we have
seen in the previous section that
for a given Lorentz factor it takes the receding
blobs apparently longer to 
travel out to the projected distance where they
stop radiating.  

Concerning the overall shape of the distribution
of the scattered radiation,
these scattering patterns appear 
more ``elongated'' along the x-axis 
for the cases where the jet axis lies close to the
plane of the sky, 
i.e.~inclination angles of roughly $90^{\circ}$.
In other words, for these cases 
more flux is contributing to the total scattered flux
from regions located at larger distances from the nucleus. 
It is interesting to compare these plots with
the corresponding plots for a point source radiating into two
oppositely directed cones as given in GSC87 for the case of 
$\theta = 90^{\circ}$ (see their Fig.~1) and by Wise \& Sarazin (1992)
for the case of $\theta = 15^{\circ}$ and also $\theta = 90^{\circ}$
(see their Fig.~7, assuming an opening angle of the cone with
a half angular width of $15^{\circ}$). 
When looking for instance at the respective
surface brightness distributions for the case where the jet axes 
lie in the plane of the sky it seems that taking into account the
actual blob motion has the effect of stretching out the whole
scattering pattern along the negative and positive x-axis,
respectively. 
Obviously, the plots provided in the earlier studies cannot provide
any brightness temperature for regions lying outside the assumed
radiation cones. However, save to the above mentioned difference 
the results inside of the radiation cones appear 
qualitatively similar.     

\begin{figure}[t]
  \centerline{\psfig{figure=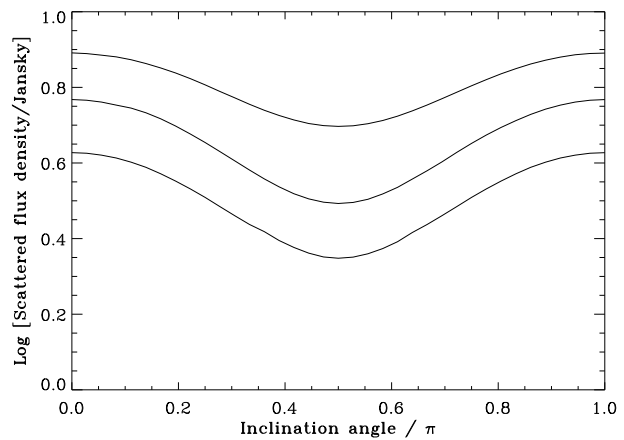,width=\columnwidth}}
 \caption{Maximum of the scattered flux density produced by a jet
          as a function
          of the inclination angle for three different Lorentz factors, 
          viz.~2, 5 and 10 (from top to bottom)
          for the assumed density distribution.}
\end{figure}

The above described behaviour is also apparent 
when looking at the dependence
of the total scattered flux on the Lorentz factor and especially
on the inclination angle. 
Fig.~12 shows the maximum of the scattered flux density produced 
by a jet as a function of the inclination angle for three different
Lorentz factors. As can be taken from this figure, 
the total scattered flux does not depend 
strongly on the Lorentz factor or the inclination angle. 

The behaviour exhibited in Fig.~12
might at first sight seem a bit surprising, in that
the dependency of the total scattered flux on the inclination 
angle and the Lorentz factor turns out to be so small.
However, since essentially the same amount of energy, 
i.e.~the same number of photons, 
is distributed over more or less the same volume,
one would expect no ``substantial'' differences between
the respective total scattered fluxes 
once the distribution 
of scattered radiation has settled to a stationary state.
We have verified that the small dependency of the
total scattered flux on the inclination angle and the Lorentz factor
apparent from Fig.~12
is primarily due to the angular dependence of the Thomson scattering 
cross section as well as to a lesser extent to the chosen density
distribution.

In Fig.~13 we are plotting the ratio of the total scattered
flux to the apparent flux for 
our simulated jet having reached its stationary state
as a function of the
inclination angle with respect to the line of sight
for three different Lorentz factors. The conclusions to be drawn from this
figure are rather similar
to the results given in Fig.~5, in that above a certain inclination angle of 
in this case about $140^{\circ}$ the total scattered
flux can dominate the apparent direct flux in the case of
a Lorentz factor of ten or larger. 
It should, however, be kept in mind that in computing the ratio
we have used the total scattered flux, to which as was shown in
the previous contour plots regions can contribute of sizes
much larger than the jet itself.

\begin{figure}[t]
  \centerline{\psfig{figure=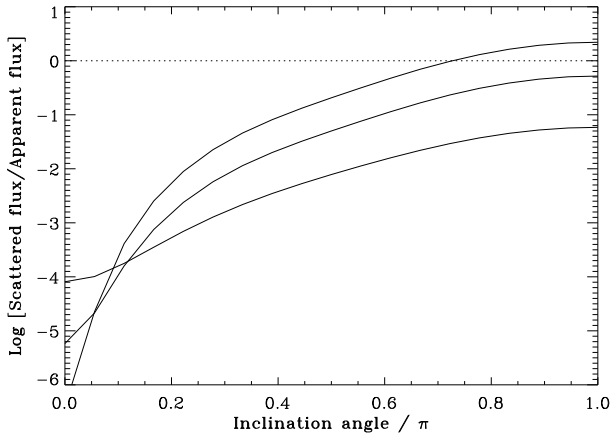,width=\columnwidth}}
 \caption{Ratio of the maximum of the scattered flux density to the 
          apparent flux for a jet as a function of 
          the inclination angle of
          the jet with respect to the 
          line of sight for three different Lorentz factors, 
          viz.~2, 5 and 10.}
\end{figure}

For Lorentz factors equal or smaller than
five the apparent direct flux dominates the total scattered
flux for any inclination angle.  
These curves do not depend on the assumed intrinsic jet luminosity,
but scale linearly with the assumed electron density. 
Since as we have seen in the previous figure the total scattered
flux varies very little with the inclination angle,
the strong dependence of the ratio given in Fig.~13
on the inclination angle essentially
is due to the variation of the beaming factor 
on the Lorentz factor and the inclination angle. 
This point will be of importance in the following section
where we try to derive constraints on the parameters 
of the jet of M87 based on the above described models.

\section{Application to M87}

\subsection{Observations}

M87 (NGC 4486) is a giant elliptical galaxy near the centre
of the Virgo cluster, which is located at a distance 
of about 16 Mpc (Tonry 1991; 1 arcsec corresponds to 78 pc).
It contains one of the best studied synchrotron jets as it is one
of the nearest examples of this phenomenon (for a general
review about the jet in M87 see e.g.~Biretta 1993).
M87 is classified as a FR I type radio galaxy, meaning a low-luminosity,
centre-brightened source.
Its approximately 2.5 kpc long jet exhibits a complex knotty structure and
is a prominent source of radio, optical and X-ray emission
(Biretta, Stern \& Harris 1991). 
The jet has a luminosity of $L_\mathrm{jet}\sim 10^{43} \, {\rm erg \,
s^{-1}}$, with the brightest knots contributing
a few times $10^{42} \, {\rm erg \, s^{-1}}$.
This non thermal activity is believed to ultimately be driven 
by a central black hole of a mass of about $3\times10^{9}\,M_{\odot}$
(e.g.~Macchetto et al.~1997).
Although a counterjet has not been observed directly, a feature 
reminiscent of a hotspot has been detected at 
optical wavelengths (Sparks et al.~1992).

\subsection{Constraints on the jet parameters}

Radio observations have allowed to derive several 
constraints on the parameters of the jet of M87. 
These constraints can be plotted in the parameter space spanned by
the Lorentz factor and the inclination angle in order 
to study which regions thereof are consistent with the observations. 
In the case of M87 such a plot has been presented 
e.g.~by Biretta
et al. (1989, 1995).
Based on these observations and the constraints derived therewith,
the inner jet (from the nucleus to the so-called knot A) is believed
to exhibit a Lorentz factor of about 3 to 5 and to be orientated at
an inclination angle of about $30^{\circ}$ to 
$40^{\circ}$ from the line of sight
(Biretta et al.~1995). 

\subsubsection{Constraint based on the flux ratio}
In deriving these constraints use was made of the fact
that the ratio of the jet flux to the counterjet flux  
\begin{equation}
R_{1}=\frac{F_{\mathrm{app}}^{\mathrm{jet}}}
{F_{\mathrm{app}}^{\mathrm{cjet}}}=
\left(\frac{1+\beta\cos\theta}{1-\beta\cos\theta}\right)^{2+\alpha}
\end{equation}
has to be at least larger than the observed lower
limit for this quantity of 150,
wherein $F_{\mathrm{app}}^{\mathrm{jet}}$ and 
$F_{\mathrm{app}}^{\mathrm{cjet}}$ are the apparent fluxes 
from the jet and the counterjet, respectively.
Obviously, Eq.~19 is only valid for the luminosity of the
jet and the counterjet being equal.
Less conservative estimates for this ratio provided in Biretta et al.~(1989)
are 740 and 1.600,
which were obtained by using a different technique of
background subtraction. These estimates
were derived by measuring the flux in a $9^{\circ}$ wide wedge 
centered on the jet up to a distance of about 19 arcsec from the
nucleus and a corresponding region located opposite of the nucleus
was used for the counterjet. 

\begin{figure}[t]
  \centerline{\psfig{figure=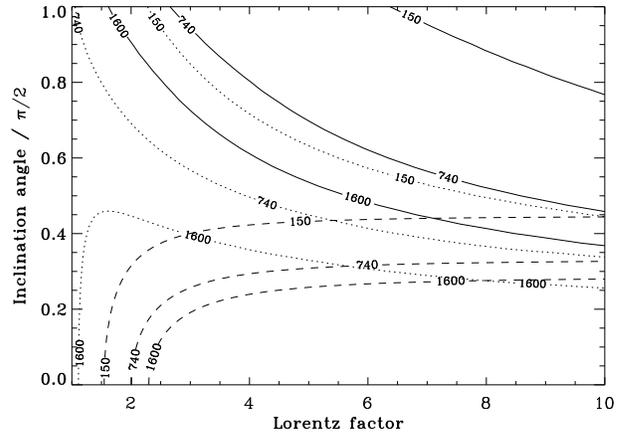,width=\columnwidth}}
 \caption{Constraints on the jet of M87 in the parameter space
          spanned by the Lorentz factor and the inclination angle.
          The solid lines mark the constraints derived by the method
          described in this
          paper based on the ratio $R_{2}$ of the apparent jet flux
          to the scattered flux received from a corresponding region 
          on the side of the counterjet having the shape of a wedge 
          with its tip centered on the nucleus.   
          The estimated lower limits for this ratio
          of 150, 740 and $1.600$ have been taken from 
          Biretta et al.~(1989). For consistency with observations
          the parameters of the jet of M87 have to lie in a region
          above these respective estimates, i.e.~``below'' the
          corresponding contours.
          The dashed lines show the conventional
          constraint based on the ratio $R_{1}$ of the apparent jet flux
          to the apparent counterjet flux.
          The dotted lines give the constraints according to 
          the present paper in case it would be possible
          to use the 
          total scattered flux.}
\end{figure}

The above cited observational lower limits for the flux ratio also set a
constraint on the scattered flux produced in the same region located 
on the side of the counterjet, in that also the ratio
\begin{equation}
R_{2}=\frac{F_{\mathrm{app}}^{\mathrm{jet}}}
{F_{\mathrm{sc}}^{\mathrm{cjet}}}
\end{equation}
has to be larger than the observational lower limits
obtained by Biretta et al.~(1989), 
where $F_{\mathrm{sc}}^{\mathrm{cjet}}$ is the scattered 
flux in the corresponding region located on the side of
the counterjet.
The model developed in the previous sections allows

us to compute the ratio $R_{2}$ for a ``grid'' of Lorentz factors
in the range from 1 to 10 and inclination angles in the range
from $0^{\circ}$ to $90^{\circ}$
using the observational parameters of M87 as follows.

Given specific values for the Lorentz factor and the inclination
angle, we set up our model to reproduce the apparent flux measured
on the side of the jet, which amounts to
about 2.4 Jy (Biretta et al.~1989). 
Furthermore, we fix the 
observed projected length of the jet to 1.5 kpc. 
We model the distribution 
of intracluster gas around M87 based on the results of
a recent analysis of XMM observations by 
Matsushita et al. (2002). The electrons are distributed
according to the sum of two beta-models (see Eq.~15)
with the following two sets of parameters:
$n_\mathrm{e}^{\mathrm{o}}=0.13\,{\rm cm}^{-3}$, 
$\beta_\mathrm{c}=0.42$, 
$r_\mathrm{c}=1.7\,{\rm kpc}$ and
$n_\mathrm{e}^{\mathrm{o}}=0.011\,{\rm cm}^{-3}$, 
$\beta_\mathrm{c}=0.47$, 
$r_\mathrm{c}=22\,{\rm kpc}$. 
We, furthermore, assume that the jet and the counterjet have
been active long enough, to reach a stationary distribution of
the scattered surface brightness as shown in Fig.~11.
Finally, we integrate this scattered surface brightness inside a 
$9^{\circ}$ wide wedge having its tip centered on the nucleus
and its symmetry axis coinciding with the axis of the counterjet
out to a distance of about 1.5 kpc from the nucleus
yielding the scattered flux in this region. 

In Fig.~14 we are plotting the two constraints 
$R_{1}$ and $R_{2}$ in the parameter space spanned
by the Lorentz factor and the inclination angle.
Obviously, the constraints derived by
the methods studied in  
this paper and based upon the 
observational limit for the jet to counterjet
flux ratio 
are less restricting than the constraints 
obtained by the conventional flux ratio method.
Or differently put, the constraint based on $R_{1}$ 
already restricts
the jet of M87 to inclination angles of less than about
$40^{\circ}$ and to Lorentz factors larger than about 1.5,
if one adopts the conservative value of 150 as a lower limit
for $R_{1}$. For the larger ratios cited in Biretta et al.~(1989)
the constraints would become correspondingly more tight.

The constraint obtained by using the ratio $R_{2}$, on the other hand,
allows us to exclude only a very small region of the parameter space
shown in Fig.~14 corresponding to Lorentz factors larger than about 8
and inclination angles larger than about $70^{\circ}$. Even
if one would adopt the larger observational limits for this
ratio, in the present case 
the constraints based on the ratio $R_{2}$ would always
be less restricting than the constraints based upon the ratio 
$R_{1}$, because the respective contour lines for the later
ratio correspondingly would exclude more and more of the 
parameter space.

One of the reasons for the constraints based upon the ratio $R_{2}$ 
being less tight than the ones 
derived based on the ratio $R_{1}$ is that in computing the
former ratio we, essentially, are missing a lot of scattered 
flux. As described above the scattered flux was computed inside 
a $9^{\circ}$ wide wedge out to a distance of about 1.5 kpc. 
However, when looking 
for instance at the snapshots presented in Fig.~11, 
one appreciates 
that a substantial fraction of the total
scattered flux originates in regions
beyond a distance of 1.5 kpc from the nucleus and also from regions
outside of the wedge in the vicinity of the nucleus. 
This ``missing'' scattered flux is also reflected in Fig.~14,
where we, furthermore, have plotted the tightest
constraints that could possibly be achieved by the method discussed
in this section. Obviously, these ``best case'' constraints,
in the present case however, 
are still less tight than the constraints based on Eq.~19
except for very large Lorentz factors of about 10.

\subsubsection{Constraint based on the brightness temperature}
A somewhat more qualitative constraint 
applying the results of the previous
sections of this paper can be obtained 
by using the observed brightness temperature
in the region of the counterjet and in the vicinity of M87.
To this end, we have searched in the literature
for appropriate observations and
have found two very interesting radio maps
of M87 and its environment 
on two different scales for the purposes of this paper.

Hines et al.~(1989) present a radio map at a wavelength
of 6 cm of the inner lobes of M87 reaching out to a distance
of about 5 kpc from the nucleus. As can be taken from Fig.~4
of their paper the surface brightness in the region of the 
counterjet lies in the range between 0.38 and 1.52 mJy per beam.
It is interesting to note that this relatively 
dim region starts very close
to the nucleus and extends out to distances of about 1.2 kpc 
therefrom along the axis of the counterjet.
Their observation was conducted at a wavelength of
6 cm and had a resolution of 0.4 arcsec. 

Concerning the distribution of the surface
brightness on larger scales Owen et al.~(2000) recently have provided
a map of the radio halo of M87 at 90 cm out to a distance 
of about 40 kpc from the nucleus. The radio map presented 
in Fig.~3 of their paper exhibits two rather interesting regions
for the purposes of the present study. These
regions lie on the respective extensions of the jet and 
the counterjet about 4 to 6 kpc distant from the nucleus
on the jet side and about 3 to 10 kpc distant from the nucleus 
on the side of the counterjet.
We estimate the surface brightness in these
relatively dim regions ``north'' of the
so-called feature D and ``south'' of the so-called feature C 
to be of the order of 20 mJy per beam (Owen et.~al 2000,
in particular Fig.~3).
The beam size of this observation was $7.8\times6.2\,{\rm arcsec}^2$.

Since the measurements of the flux obtained by Biretta et al.~(1989) 
and of the surface brightnesses in the above cited observations
are given at different wavelengths, viz.~2, 6 and 90
centimeters, one needs an estimate of the spectral 
index in order to compare these observations
at different wavelengths. 
Harris et al.~(2000) assumed spectral index values
of 0.8 and 0.9 to integrate the radio spectrum
between
$10^{7}$ and $10^{10}$
in a study comparing
extra nuclear X-ray and radio features in M87.
The values for the spectral index were obtained
from an unpublished spectral index map
between 74 and 327 MHz provided by N.~Kassim.
In a study of the radio lobes of Virgo A at 2.8 cm
wavelength Rottmann et al.~(1996) find a very steep spectrum in
the outer lobes. Values for the spectral index between
1.6 and 2.8  are quoted.
Adopting a spectral index of 1,
which should represent a conservative choice, we compute
the brightness temperatures corresponding to the limits
given by Hines et al.~(1989) and Owen et al.~(2000) to
be 25 and 0.07 Kelvin at a wavelength
of 2 cm, respectively.
Choosing a respective spectral index of 0.5 or 2 instead,
would yield respective limits for the brightness temperature
of 43 and 0.48 Kelvin as well as 8.3 and 0.002 Kelvin.

Given the above derived observational limits on the surface
brightness at a wavelength of 2 cm in two specific regions in the 
vicinity of M87, we are able by using the methods presented in this paper
to address the question, whether these observational limits
are consistent with the scattered surface brightness
to be expected from a jet producing the observed direct flux
as measured by Biretta et al.~(1989). 
Thus, in computing the respective surface brightness
profiles we proceed in a similar way to the previous 
section while taking the beam sizes of the respective
observation into account, i.e.~we fix the observed apparent 
jet flux as well as the observed projected jet length and vary the Lorentz
factor and the inclination angle. 

\begin{figure}[t]
  \centerline{\psfig{figure=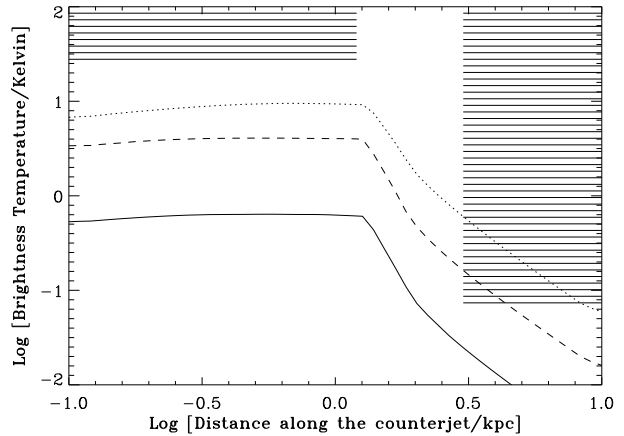,width=\columnwidth}}
 \caption{Brightness temperature of the scattered radiation
          as a function
          of the distance from the nucleus along the axis of the
          counterjet for the following
          combinations of Lorentz factor and inclination angle:
          5, $30^{\circ}$ (solid curve); 10, $30^{\circ}$ (dashed
          curve) and 5, $60^{\circ}$ (dotted curve). The observational
          upper limits marked by rectangles filled by thick horizontal
          lines are taken from Hines et al.~(1989) 
          for distances of 0.1 up to 1.2 kpc
          from the nucleus and from Owen et al.~(2000) for
          distances from 3 kpc up to 10 kpc from the nucleus.}
\end{figure}

Fig.~15 shows the results of these computations 
in that
the brightness temperature of the scattered radiation 
is plotted as a function
of the distance from the nucleus 
along the axis of the counterjet 
for three different
combinations of Lorentz factor and inclination angle, 
viz.~$\gamma=5$ and $\theta=30^{\circ}$, 
$\gamma=10$ and $\theta=30^{\circ}$ as well as
$\gamma=5$ and $\theta=60^{\circ}$. 
Also shown are the observational
upper limits taken as outlined above 
from Hines et al.~(1989) for 
distances from 0.1 up to 1.2 kpc
from the nucleus and from Owen et al.~(2000) for
distances from 3 kpc up to 10 kpc from the nucleus.

As can be taken from this figure, 
the presently most favoured model 
for the jet of M87 having 
a Lorentz factor between 3 and 5
and an inclination angle between
$30^{\circ}$ and $40^{\circ}$ lies 
well within our constraints.
However, the surface brightness profiles 
for the other two combinations of Lorentz factor
and inclination angle, i.e.~$\gamma=10$ 
and $\theta=30^{\circ}$ as well as
$\gamma=5$ and $\theta=60^{\circ}$, do not
seem to be consistent with the observations of 
Owen et al.~(2000), in that the scattered
brightness temperatures are larger than the 
ones observed. 
Obviously, for even larger Lorentz factors and/or
inclination angles this inconsistency would become
even more striking.
Again, it should be noted, that the
conclusions to be drawn from Fig.~15
are based on the assumption that the counterjet is 
intrinsically symmetric to the jet and that the 
jet has been active long enough for the scattered
radiation to reach a stationary state as described
in the previous sections.

The surface brightness level
derived from Owen et al.~(2000) represents a fundamental
restriction for our method which cannot be overcome by
observations with a very high dynamic range, since 
nowhere inside the large scale lobes of the order of about
40 kpc a lower surface brightness can be found.
Furthermore, it should be noted that several features in the
lobe on the side of the counterjet, i.e.~the east
lobe, are quite luminous (Hines et al.~1989).
The total flux of these sources is about 1.5 Jy.
Assuming this radiation to be isotropic the
scattered flux should be about 1.5 mJy for the
assumed density distribution of M87. The regions studied 
in this section are located within a circle of at least 
1 kpc around this lobe. Assuming that this scattered
flux is distributed over such an area in the sky,
a conservative upper value for the average brightness 
temperature of the scattered radiation produced by the
features within this eastern radio lobe turns out to be 
0.05 Kelvin and thus should not be critical for the
method studied here.  

For the purposes of this paper it would be very interesting
to obtain an estimate of the scattered surface brightness
at higher frequencies in the region in the vicinity of M87
we have been focusing on in this section.
This is because the 
overall synchrotron spectrum of the jet 
of M87 is
characterized by a medium flat radio to optical spectral
index of 0.65, a steep cutoff at frequencies beyond a critical
value of about $2\times 10^{15}$ Hz and some flattening
at frequencies below $10^{10}$ Hz
(Meisenheimer et al.~1996). Given the above mentioned 
steep spectrum of the outer radio lobes, wherein this
region is located, observations at 
higher frequencies, therefore, might improve the
results presented here.

\section{Summary}
In this paper we have tried to assess the possibility
of constraining the parameters of relativistic
extragalactic jets by means of a model for the radiation 
field produced by such jets. Based on this model we have 
computed 
the brightness temperature of the scattered radiation
as well as the scattered fluxes for several different
scenarios.  

In a first part of this paper we have illustrated and 
discussed the scattered radiation produced by a single 
pair of blobs being ejected along oppositely directed 
axes out of the nucleus of an AGN. In a second part we 
have studied the distribution of the
scattered surface brightness produced by multiple blob pairs.

The results of these two sections have been used in a final section 
of this paper, where we have tried to derive independent 
constraints for the jet of M87 based on the observed
jet to counterjet flux ratio and based on the surface
brightness in two specific regions on the side of the
counterjet. We found that the presently most popular
model for the jet of M87 lies well within the derived constraints.
However, a combination of a very large Lorentz factor and a 
large inclination angle does not seem to be 
consistent with the observational limits.

Lorentz factors of up to 10 have been detected in extragalactic jets. 
As is apparent from the foregoing results, any
extragalactic jet located in a cluster with a high electron density
might be a promising target for the method presented in this paper.
However, motions with apparent superluminal velocities have been
observed in galactic objects as well.
For instance the two microquasars GRS 1915+105 and GRO J1655-40
(Mirabel \& Rodriguez 1994, Hjellming \& Rupen 1995) 
have an inferred bulk Lorentz factor of 2.5.
However, the observations seem to suggest that 
the single blob scenario should be more appropriate
for these cases.

Observations 
of Gamma-Ray Bursts (GRBs)
strongly indicate and theoretical models 
for these objects advocate,
the presence of relativistic, well collimated outflows in GRBs.
The radiating matter is believed to move initially 
with Lorentz factors up to $10^2-10^3$.
From the perspective of the present paper these objects are very
interesting in that the inclination angle is most probably 
very well constrained, since
these objects are believed to be observed 
very close to the line of sight.
In this context it should be stressed, that in principle 
the method studied in this paper
to constrain relativistic, well collimated outflows 
is not restricted to radio wavelengths.
Studies in this direction have been presented 
by Sazonov \& Sunyaev (2003), who have demonstrated
how the observed X-ray radiation from GRBs can provide
information about the collimation angles of these objects.   

\begin{acknowledgements}
We thank 
E.~Churazov and M.~Gilfanov
for enlightening discussions.
\end{acknowledgements}

\end{document}